УДК 681.391+511.9+519.612

# ОПРЕДЕЛЕНИЕ КОДОВОГО РАССТОЯНИЯ НЕДВОИЧНОГО LDPC-КОДА БЛОЧНЫМ МЕТОДОМ КОРКИНА-ЗОЛОТАРЕВА

*В.С. Усатюк, аспирант кафедры вычислительной техники, ЮЗГУ (e-mail:L @Lcrypto.com)*

В данной работе методы геометрии чисел: блочный метод Коркина-Золотарева (приведение базиса решетки) и метод Каннана-Финке-Поста (поиск кратчайшего вектора) были применены для решения задач оценки кодовых расстояний помехоустойчивых линейных блочных кодов. В работе предложен метод для оценки кодовых расстояний тернарных кодов GF(3), с целью выбора наилучших LDPC-кодов для их отображений на гексагональное созвездие. Этот метод также применим для оценки кодовых расстояний двоичных образов недвоичных GF(64) LDPC-кодов с целью выбора кода для двоичной и квадратурной фазовых модуляций (BPSK, QPSK), применяемых при оптической передаче информации и в архивной голографической памяти. В работе получена верхняя оценка константы масштабирующего коэффициэнта в случае приведения базиса блочным методом Коркина-Золотарева с размером блока $\beta$.

Ключевые слова: кодовое расстояние, LDPC-код, приведение базиса решетки, целочисленные решетки, поиск кратчайшего базиса, поиск кратчайшего вектора, блочный метод Коркина-Золотарева.

## 1. Введение

Впервые двоичные LDPC-коды были предложены Робертом Галлагером в его докторской диссертации 1963 года [1]. Однако эти результаты на протяжении более 30 лет были проигнорированы. Причинами этого являлись отсутствие строгих аналитических методов поиска хороших LDPC-кодов и сравнительно высокая сложность алгоритмов их декодирования.

Первый алгебраический метод построения двоичных LDPC-кодов был предложен Маргулисом [2]. Однако из-за псевдокодовых слов минимального веса, обусловленных пересечениями циклов графа при декодировании методом распространения доверия, фактическое кодовое расстояние получилось равное 24 вместо теоретически предсказанного 220 [3]. Код Маргулиса продемонстрировал слабость существующих в то время алгебраических методов помехоустойчивого кодирования при решении задач построения кодов для мягкого декодирования методом распространения доверия.

Более поздняя работа Таннера продемонстрировала новый теоретико-графический подход и связь двух ключевых параметров LDPC-кодов,

кодового расстояния и обхвата графа. [4]. Осмысление работы Таннера привело к возникновению широкого класса алгебраических методов построения LDPC-кодов. Были предложены: Евклидово и проективно-геометрические LDPC-коды (EG/PG, [5]); LDPC-коды на словах Рида-Соломона (RS-LDPC, [6]); коды на блочных неполных дизайнах (BIBD, [7]) и других структурах инцидентности, принадлежащих классу графов Рамануджана [8]. Ключевым достоинством алгебраических методов является их конструктивная природа, знание кодового расстояния, обхвата, скорости кода, автоморфизма графа и других свойств. Главным недостатком алгебраических методов является сложность изменения их параметров, не допускающая получения кодов с гибким диапазоном скоростей, длин, строчных и столбчатых весов, с увеличенным обхватом графа без использования масок.

По этой причине параллельно с алгебраическими методами бурно развивались случайные методы построения кодов на графах: Guess-and-test ([9]), Bit-Filing ([10]), PEG ([11]), Hill-climbing ([12]). Однако все они требуют оценки ключевых параметров кода, в частности кодового расстояния.

Недвоичные LDPC-коды впервые были предложены в работе Д. Маккея, М. Дэви ([13]). Маккеем был предложен эвристический метод построения таких кодов при помощи энтропии синдрома [14]. В работах Фоссорье и Деклерка был изложен метод оптимизации назначения недвоичного значения через оптимизацию спектра строчных кодов и разрыв циклов [15].

Значительный энергетический выигрыш помехоустойчивые коды обеспечивают при их согласовании с методами модуляции. В работе Зигангирова К.Ш., Енгдала К. был продемонстрирован энергетический выигрыш в случае применения гексагонального созвездия (пять трит, $3^5=243$) в сравнении с прямоугольной многоуровневой квадратурной модуляцией (восемь бит, $2^8=256$, QAM-256) [16]. В работе была продемонстрирована эффективность использования геометрических решеток (Lattice)-наиплотнейших упаковок в качестве модуляционных созвездий, обуславливающих выигрыш за счет формы созвездия ("shape gain"), ранее описанный в пионерских работах Вея-Форни [17], Полтырева [18], Де Будде [19] и Урбанке [20].

Для согласования созвездия и кода требуется получение тернарных кодов с наибольшим кодовым расстоянием. Существуют методы построения совершенных тернарных кодов ([20]), однако для случая произвольного тернарного кода, даже для малых длин задача не была решена, в том числе и численными методами поиска кодового расстояния [21].

В данной работе предлагается метод определения кодового расстояния тернарного LDPC-кода с помощью блочного метода Коркина-Золотарева. В

отличие от импульсного метода Берроу К. [22] и метода Фоссорье-Ху-Эльфеора [23] поиска слов малого веса методом распространения доверия, применимого к кодам, допускающим мягкое декодирование: Турбо-коду и LDPC-кодам, предложенный метод применим к произвольному линейному двоичному и тернарному кодам. Кроме того метод дает оценку истинного кодового расстояния, а не оценку веса псевдокодового слова.

Согласование прямоугольной модуляции [24] с недвоичным кодом также позволяет получить дополнительный энергетический выигрыш. По этой причине предложенный метод модифицируется для оценки кодового расстояния двоичного образа недвоичного GF(q) LDPC-кода (q>3) в случае использования двоичной и квадратурной фазовых модуляций (BPSK , QPSK), [25].

## 2. Основные понятия и обозначения

**Определение 1.** Решетка - дискретная абелева подгруппа, заданная в пространстве $R^n$.

Пусть базис $B = \{b_1,...,b_n\}$ задан линейно-независимыми векторами в $R^m$. Тогда под решеткой будем понимать множество целочисленных линейных комбинаций этих векторов:

$$L(b_1,...,b_n) = \{\sum_{i=1}^{n} x_i b_i : (x_1,...,x_n) \in Z^n\},$$

где $m$ и $n$, размерность и ранг решетки соответственно, $m \geq n$.

**Определение 2.** Задача поиска короткого вектора ($\Delta$-short vector problem, $SVP_\Delta(m)$): Пусть дана $m$-мерная решетка $L(B)$, ранга $n$ и вещественное $\Delta > 1$. Найти нетривиальный вектор в $\Delta$-раз больший кратчайшего вектора в решетке $\bar{b} \in L : \|\bar{b}\| \leq \Delta \cdot \lambda_1(L)$.

При $\Delta = 1$, решается задача поиска кратчайшего вектора в решетке, при $\Delta > 1$ - короткого вектора.

**Определение 3.** Задача поиска короткого базиса решетки ($\Delta$-short basis problem, $SBP_\Delta(m)$): Пусть дан базис полной решетки $B$ и вещественное $\Delta > 1$. Найти базис $B' = \{b_1', b_2',...,b_m'\} : L(B) = L(B'), \prod_{i=1}^{m} \|b_i'\| \leq \Delta \cdot \prod_{i=1}^{m} \|b_i^\perp\|$.

Подробнее методы геометрии чисел: алгоритм поиска короткого базиса решетки (блочный метод Коркина-Золотарева) и алгоритм поиска кратчайшего вектора рассмотрены в работах [26].

**Определение 4.** Линейный код длины $n$ и ранга $k$ является линейным подпространством $C$ размерности $k$ пространства $F_q^n$, где $F_q$ - конечное поле из $q$ элементов (символов).

**Определение 5.** Линейный блоковый код – такой код, что множество его кодовых слов образует $k$-мерное линейное подпространство $C$ в $n$-мерном линейном пространстве, изоморфное пространству $k$-символьных векторов.

Операция кодирования осуществляется умножением исходного $k$-символьного вектора на невырожденую матрицу $G$, называемую порождающей матрицей линейного блочного кода.

В данной статье мы будем расматривать линейные блоковые тернарные LDPC-коды ($q=3$) и двоичные образы 64-арного LPDC-кода.

**Определение 6.** Расстоянием Хемминга между двумя словами $\bar{v}_1$ и $\bar{v}_2$ называется количество символов, в которых слова различаются.

**Определение 7.** Минимальное расстояние $d_{\min}$ линейного кода является минимальным из всех расстояний Хемминга всех пар кодовых слов.

**Определение 8.** Вес кодового слова $\omega$ равен расстоянию Хемминга между этим кодовым словом и нулевым кодовым словом. Вес $\omega$ определяется как число ненулевых компонент кодового слова.

**Определение 9.** Весовой спектр кода это вектор $\omega(C) = (w_0, w_1, w_2, ..., w_i, ..., w_n)$, где $w_i$ - количество кодовых слов веса $i$.

**Определение 10.** Двудольный граф Таннера — это граф, множество вершин которого можно разбить на две части, соответствующие символьным и проверочным узлам, таким образом, что каждое ребро графа соединяет какую-то вершину из одной части с какой-то вершиной другой части. Не существует ребра, соединяющего две вершины из одной и той же части.

**Определение 11.** Обхват графа — длина наименьшего цикла, содержащегося в данном графе.

**Определение 12.** Величина автоморфизма графа — число различных нетривиальных изоморфизмов графа на себя, отображений сохраняющих отношение смежности между вершинами графа.

**Определение 13.** Базис $B = \{b_1, b_2, ..., b_m\}$ решётки $L \subset R^m$ приведён по длине, если в результате ортогонализации решетки методом Грамма-Шмидта выполняется следующее неравенство:
$$|\mu_{i,j}| \leq \frac{1}{2}, 1 \leq j < i \leq m,$$
где $\mu_{i,j}$ - коэффициенты Грамма-Шмидта.

**Определение 14.** Упорядоченный по длине базис $B = \{b_1, b_2, \ldots, b_m\}$ решётки $L \subset R^m$ приведён блочным методом Коркина-Золотарева с блоком $\beta \in [2, m]$ и точностью $\delta \in \left(\frac{1}{2}; 1\right]$, если:

- базис $B$ приведён по длине;
- $\delta^2 \cdot \left\|b_i^\perp\right\|^2 \leq \lambda_1^2(L_i), i = 1, \ldots, m,$ где $\lambda_1(L_i)$ - длина кратчайшего вектора в решётке $L_i$, образованной ортогональным дополнением векторного пространства с базисом $b_i, \ldots, b_{\min(i+\beta-1, m)}$.

**Определение 15.** Базис решетки приведен по Ленстра-Ленстра-Ловасу (LLL-алгоритмом), если он приведен блочным методом Коркина-Золотарева с размером блока $\beta = 2$ и точностью ортогонализации $\delta \in \left(\frac{1}{2}; 1\right)$.

### 3. Метод определения кодового расстояния линейного блочного кода с использованием блочного метода Коркина-Золотарева

Для определения кодового расстояния нам необходимо найти в коде слова малого веса. С этой целью нам необходимо вложить код в решетку - отмасштабировать $n - k$ подпространств базиса решетки $B_c$ некоторой константой $N$ такой, чтобы расстояние Хэмминга между кодовыми словами отобразилось в Евклидово расстояния между векторами кодовых слов решетки. Легко убедится, что полученная решетка – вырожденная, ранга $rank(B_c) = k$.

Осуществив поиск короткого базиса решетки методом Коркина-Золотарева, получим некоторый набор векторов малого веса с нормой $r_{\max}$. Вектор решетки с наименьшей нормой соответствует предполагаемому кратчайшему вектору. Однако пока мы не осуществим перебор всех линейных комбинаций методом Каннана-Финке-Поста и не убедимся, что полученная норма вектора является наименьшей возможной в решетке, данное кодовое слово будем считать словом малого веса.

Оценим масштабирующий коэффициент $N$ необходимый для корректной работы метода:

**Теорема 1.** Базис решетки $L(B_c)$ несистематического кода, $\delta$-приведённый по Ленстра-Ленстра-Ловасу, в качестве приведенных коротких базисных векторов решетки будет давать слова малого веса несистематического кода, если константа:

$$N \geq \left(\frac{4}{4\delta - 1}\right)^{n/2} \times 2\sqrt{(m+1)(n+1)} \times r_{\max} \times M^m,$$

где $M = \max\{\|A_0\|, \|A_1\|, \ldots, \|A_{n-1}\|, \|d\|\}$ эквивалентного линейного диофантова уравнения $A \times x = d$.

Доказательством теоремы приведено в работе [27, стр. 594-596]. Оценим $N$ в случае приведения базиса блочным методом Коркина-Золотарева с размером блока $\beta$.

**Теорема 2.** Базис решетки $L(B_c)$ несистематического кода, приведённый блочным методом Коркина-Золотарева с размером блока $\beta$, в качестве приведенных коротких базисных векторов решетки будет давать слова малого веса несистематического кода, если константа:

$$N \geq \gamma_\beta^{\frac{n-1}{\beta-1}} \times 2\sqrt{(m+1)(n+1)} \times r_{\max} \times M^m.$$

**Доказательство**:

Первая часть доказательства берется из доказательства Теоремы 1. Длина векторов, образующих избыточности в исходном коде, ограничена сверху величиной $\|V_i\| \leq 2\sqrt{(n+1)(m+1)} \times r_{\max} \times M^m, 0 \leq i \leq n-k$.

Для доказательства оставшейся части нам необходимо оценить сверху длину векторов избыточности решетки после приведения ее базиса блочным методом Коркина-Золотарева с размером блока $\beta$, $\overline{b_i} \in L': 0 \leq i \leq n-k$. В работе [28] Шнор доказал, что длина приведенных векторов ограничена сверху величиной $\|b_i\|^2 \leq \gamma_\beta^{2\frac{n-1}{\beta-1}} \frac{i+3}{4} \lambda_i^2$, где $\gamma_\beta$ -константа Эрмита, $\lambda_i$ - $i$ – соответствующий минимум решетки. Подстановкой получим искомую оценку.

В случае систематической порождающей матрицы кода, легко убедиться, что константа $N = 1$.

Предлагаемый метод поиска кодовых слов минимального веса требует построения решетки на основе порождающей матрицы систематического кода $G = (I_k \mid A), A \in F_q^{k \times n-k}$

$$B_c = \begin{pmatrix} A^T & qI_{n-k} \\ I_k & 0 \end{pmatrix}, (1)$$

В случае несистематического кода:

$$B_c = \begin{pmatrix} N \cdot G' & N \cdot qI_{n-k} \\ I_k & 0 \end{pmatrix}, (2)$$

где $G'$ - порождающая матрица несистематического кода, $q$ - количество символов, $I_k$ - единичная матрица [31]

После чего в решетке ищется вектор $\|v\|_\infty = 1, v \in L, rank(L) = n$ с минимальным числом отличных от нуля координатных компонент. Переход от метрики Евклида к метрики Чебышева дает нам эквивалентную формулировку задачи поиска слов минимального веса в случае решетки, а именно поиск среди точек решетки, образованных различными $2^n$ шарами с центрами $\left(\pm\dfrac{d}{2},...,\pm\dfrac{d}{2}\right)$ и радиусами $\dfrac{d}{2}\sqrt{n}$ [29]. Кодовые слова двоичного $q = 2$, $-1 \equiv 1 \bmod 2$ и троичного кодов $q = 3$, $-1 \equiv 2 \bmod 3$ будут представлены векторами решетки, с координатными компонентами, принимающими значения из множества $\{-1, 0, 1\}$.

Формальное описание метода представлено ниже в виде алгоритма:

Вход алгоритма: порождающая матрица кода, вложенная в решетку $L \in R^{m+n+1}$ ранга $rank(L) = n+1$ с базисом, заданным формулами 1 и 2, в случае систематического и несистематического кода, соответственно.

1. Приведем решетку блочным метод Коркина-Золотарева с размером блока $\beta$;

2. После удаления линейно зависимых строк и столбцов получим некоторый короткий базис решетки $\bar{b}_0, \bar{b}_1, ..., \bar{b}_{n-m} \in R^{n+1}$, $rank(L(b_0, b_1, ..., b_{n-m})) = n - m + 1$. В случае, если среди получившихся коротких векторов, будет вектор нормы меньше изначальной оценки, примем длину полученного вектора, соответсвующего некоторому слову малого веса, за новую оценку с целью уменьшения дерева перебора;

3. Ортогонализуем полученный базис методом Грамма-Шмидта, получим ортогональное дополнение пространства решетки $\bar{b}_0^\perp, \bar{b}_1^\perp, ..., \bar{b}_{n-m}$ и коэффициенты Грамма-Шмидта $\mu_{ij}$;

4. Выполним полным перебором поиск кратчайшего вектора $\bar{v} \in Z^{n+1} : -r_{\max} \leq v_i \leq -r_{\max}, i = 0..n$ (алгоритм Каннана-Финке-Поста) в решетки с начальной нормой $A = (n+1) \times r_{\max}^2$.

На выходе алгоритма получается кодовое слово минимального веса Хэмминга.

На шаге 3 с целью распараллеливания метода ортогонализации можно применить преобразование Хаусхолдера [26]. Шаг 4 распараллеливается аналогичным способом.

Рассмотрим работу алгоритма на примере тернарного совершенного кода Голея (11,6), с порождающей матрицей:

$$G = \begin{pmatrix} 2 & 2 & 1 & 2 & 0 & 1 & 0 & 0 & 0 & 0 & 0 \\ 0 & 2 & 2 & 1 & 2 & 0 & 1 & 0 & 0 & 0 & 0 \\ 0 & 0 & 2 & 2 & 1 & 2 & 0 & 1 & 0 & 0 & 0 \\ 0 & 0 & 0 & 2 & 2 & 1 & 2 & 0 & 1 & 0 & 0 \\ 0 & 0 & 0 & 0 & 2 & 2 & 1 & 2 & 0 & 1 & 0 \\ 0 & 0 & 0 & 0 & 0 & 2 & 2 & 1 & 2 & 0 & 1 \end{pmatrix}$$

Решетка для поиска минимального расстояния в случае приведения блочным методом Коркина-Золотарева с размером блока $\beta = 2$ и N=6 примет вид:

$$B_c^T = \begin{pmatrix} 12 & 12 & 6 & 12 & 0 & 6 & 0 & 0 & 0 & 0 & 0 & 1 & 0 & 0 & 0 & 0 & 0 \\ 0 & 12 & 12 & 6 & 12 & 0 & 6 & 0 & 0 & 0 & 0 & 0 & 1 & 0 & 0 & 0 & 0 \\ 0 & 0 & 12 & 12 & 6 & 12 & 0 & 6 & 0 & 0 & 0 & 0 & 0 & 1 & 0 & 0 & 0 \\ 0 & 0 & 0 & 12 & 12 & 6 & 12 & 0 & 6 & 0 & 0 & 0 & 0 & 0 & 1 & 0 & 0 \\ 0 & 0 & 0 & 0 & 12 & 12 & 6 & 12 & 0 & 6 & 0 & 0 & 0 & 0 & 0 & 1 & 0 \\ 0 & 0 & 0 & 0 & 0 & 12 & 12 & 6 & 12 & 0 & 6 & 0 & 0 & 0 & 0 & 0 & 1 \\ 18 & 0 & 0 & 0 & 0 & 0 & 0 & 0 & 0 & 0 & 0 & 0 & 0 & 0 & 0 & 0 & 0 \\ 0 & 18 & 0 & 0 & 0 & 0 & 0 & 0 & 0 & 0 & 0 & 0 & 0 & 0 & 0 & 0 & 0 \\ 0 & 0 & 18 & 0 & 0 & 0 & 0 & 0 & 0 & 0 & 0 & 0 & 0 & 0 & 0 & 0 & 0 \\ 0 & 0 & 0 & 18 & 0 & 0 & 0 & 0 & 0 & 0 & 0 & 0 & 0 & 0 & 0 & 0 & 0 \\ 0 & 0 & 0 & 0 & 18 & 0 & 0 & 0 & 0 & 0 & 0 & 0 & 0 & 0 & 0 & 0 & 0 \\ 0 & 0 & 0 & 0 & 0 & 18 & 0 & 0 & 0 & 0 & 0 & 0 & 0 & 0 & 0 & 0 & 0 \\ 0 & 0 & 0 & 0 & 0 & 0 & 18 & 0 & 0 & 0 & 0 & 0 & 0 & 0 & 0 & 0 & 0 \\ 0 & 0 & 0 & 0 & 0 & 0 & 0 & 18 & 0 & 0 & 0 & 0 & 0 & 0 & 0 & 0 & 0 \\ 0 & 0 & 0 & 0 & 0 & 0 & 0 & 0 & 18 & 0 & 0 & 0 & 0 & 0 & 0 & 0 & 0 \\ 0 & 0 & 0 & 0 & 0 & 0 & 0 & 0 & 0 & 18 & 0 & 0 & 0 & 0 & 0 & 0 & 0 \\ 0 & 0 & 0 & 0 & 0 & 0 & 0 & 0 & 0 & 0 & 18 & 0 & 0 & 0 & 0 & 0 & 0 \end{pmatrix}.$$

После приведения базиса решетки получим:

$$B_c = \begin{pmatrix} 0 & 0 & 0 & 0 & 0 & 0 & 0 & 0 & 0 & 0 & 0 & 0 & -3 & 0 & 0 & 0 & 0 \\ 0 & 0 & 0 & 0 & 0 & 0 & 0 & 0 & 0 & 0 & 0 & 0 & 0 & -3 & 0 & 0 & 0 \\ 0 & 0 & 0 & 0 & 0 & 0 & 0 & 0 & 0 & 0 & 0 & 0 & 0 & 0 & 0 & 0 & -3 \\ 0 & 0 & 0 & 0 & 0 & 0 & 0 & 0 & 0 & 0 & 0 & -3 & 0 & 0 & 0 & 0 & 0 \\ 0 & 0 & 0 & 0 & 0 & 0 & 0 & 0 & 0 & 0 & 0 & 0 & 0 & 0 & -3 & 0 & 0 \\ 0 & 0 & 0 & 0 & 0 & 0 & 0 & 0 & 0 & 0 & 0 & 0 & 0 & 0 & 0 & -3 & 0 \\ 6 & 6 & 6 & 0 & 0 & 0 & 6 & 0 & 0 & 6 & 0 & -1 & 0 & 1 & 0 & 1 & 0 \\ 6 & 0 & 0 & -6 & 0 & -6 & 0 & 0 & -6 & 6 & 0 & -1 & 1 & 1 & -1 & 1 & 0 \\ 0 & 6 & 6 & 6 & 0 & 0 & 0 & 6 & 0 & 0 & 6 & 0 & -1 & 0 & 1 & 0 & 1 \\ 6 & 0 & 0 & 0 & 6 & 0 & 0 & 6 & 0 & 6 & 6 & -1 & 1 & 1 & 1 & 1 & 1 \\ 6 & 6 & 0 & -6 & -6 & 0 & 0 & -6 & 0 & 0 & 0 & -1 & 0 & -1 & 0 & 0 & 0 \\ 0 & -6 & 0 & 6 & 0 & 6 & -6 & 0 & 0 & 0 & 6 & 0 & 1 & -1 & 1 & 0 & 1 \\ 0 & -6 & 0 & 0 & 0 & 0 & 0 & 6 & 6 & -6 & 6 & 0 & 1 & -1 & -1 & -1 & 1 \\ -6 & 0 & -6 & -6 & -6 & 0 & 0 & 0 & -6 & 0 & 0 & 1 & -1 & 0 & -1 & 0 & 0 \\ 0 & 0 & -6 & 0 & -6 & -6 & -6 & 0 & 0 & 0 & -6 & 0 & 0 & 1 & -1 & 0 & -1 \\ 0 & 0 & 0 & -6 & 0 & 0 & -6 & 0 & -6 & -6 & -6 & 0 & 0 & 0 & 1 & -1 & -1 \\ 0 & 0 & -6 & 0 & 0 & -6 & 0 & -6 & -6 & -6 & 0 & 0 & 0 & 1 & -1 & -1 & 0 \end{pmatrix}.$$

Отбрасываем первые 6 строк и последние 6 столбцов, как вспомогательные координатные компоненты (подпространство векторов избыточности кода), и отмасштабируем значения координатных компонент в решетке. Получим базис решетки, дающий некоторое слово малого веса, которое может оказать словом наименьшего веса:

$$B_c = \begin{pmatrix} 1 & 1 & 1 & 0 & 0 & 0 & 1 & 0 & 0 & 1 & 0 \\ 1 & 0 & 0 & -1 & 0 & -1 & 0 & 0 & -1 & 1 & 0 \\ 0 & 1 & 1 & 1 & 0 & 0 & 0 & 1 & 0 & 0 & 1 \\ 1 & 0 & 0 & 0 & 1 & 0 & 0 & 1 & 0 & 1 & 1 \\ 1 & 1 & 0 & -1 & -1 & 0 & 0 & -1 & 0 & 0 & 0 \\ 0 & -1 & 0 & 1 & 0 & 1 & -1 & 0 & 0 & 0 & 1 \\ 0 & -1 & 0 & 0 & 0 & 0 & 0 & 1 & 1 & -1 & 1 \\ -1 & 0 & -1 & -1 & -1 & 0 & 0 & 0 & -1 & 0 & 0 \\ 0 & 0 & -1 & 0 & -1 & -1 & -1 & 0 & 0 & 0 & -1 \\ 0 & 0 & 0 & -1 & 0 & 0 & -1 & 0 & -1 & -1 & -1 \\ 0 & 0 & -1 & 0 & 0 & -1 & 0 & -1 & -1 & -1 & 0 \end{pmatrix}$$

В качестве базиса решетки мы получили одиннадцать векторов, соответсвующих 11 кодовым словам веса 5. Вектору решетки в последней строке $v = (0,0,-1,0,0,-1,0,-1,-1,-1,0)$ соответствует кодовое слово $(0,0,2,0,0,2,0,2,2,2,0)$ веса 5. Для того, чтобы убедиться, что не существуют слова меньшего веса, нам необходимо выполнить полный перебор среди всех возможных линейных комбинаций векторов решетки, отличающихся своими координатными компонентами. Запуск алгоритма Каннна-Финке-Поста показывает, что таковые вектора в решетке отсутствуют, $d_{\min} = 5$.

Корректность представленного алгоритма и программного комплекса его осуществляющего были проверены на более чем 30 различных кодах. Использовались Полярные коды, коды Рида-Маллера, коды БЧХ, код Голея, Турбо-коды, LDPC-коды [30]. Например, для оценки кодового расстояния эквивалентной блочной матрицы Турбо-кода [156, 48,13] при однопоточной реализации метода на ЭВМ (Phenom x4-965/ 8 Gb DDR3) потребовалась 21 секунда, тогда как оценка кодового расстояния в GAP 4.7.8 (Guava 3.12, Sonata 2.6) требует более 648000 секунд. Однопоточная реализация предложеного метода доступна на сайте автора [31].

**5. Результаты оценки кодового расстояния недвоичных LDPC-кодов с помощью предложенного метода**

С использованием предложенного метода нами были построены 5 квазициклических (автоморфизм графа равен размеру циркулянта) регулярных тернарных LDPC-кодов с параметрами, приведенными в таблице 1:

Таблица 1

Кодовые растояние тернарных LDPC-кодов

| Длина кода | Вес столбца | Вес строки | Циркулянт | Обхват | $d_{\min}$ |
|---|---|---|---|---|---|
| 300 | 3 | 6 | 50 | 8 | 59* |
| 560 | 4 | 8 | 70 | 8 | 28 |
| 564 | 4 | 12 | 47 | 6 | 12 |
| 940 | 4 | 20 | 47 | 6 | 15 |
| 2400 | 4 | 40 | 60 | 6 | 12 |

*-после исследования 178 из 300 измерений решетки

Для построения оптимизированных кодов мы использовали 4 проверочных матрицы древоподобных кодов (с минимальным автоморфизмом графа), построенных Деклерком при помощи PEG-алгоритма [14,32]. На основе метода назначения недвоичного значения на ребре [15] нами были построены три квазициклических кода. Результаты оценки кодового расстояния двоичного образов этих кодов представлены в Таблице 2:

Таблица 2

Кодовые растояние двоичных образов GF(64) LDPC-кодов

| Длина кода | Вес столбца | Вес строки | Циркулянт | Обхват | $d_{\min}$ |
|---|---|---|---|---|---|
| 52 | 2 | 4 | - | 12 | 6 |
| 64 | 2 | 8 | - | 8 | 4 |
| 210 | 2 | 20 | - | 6 | 4 |
| 820 | 2 | 40 | - | 6 | 4 |
| 212 | 2 | 4 | 53 | 12 | 6 |
| 240 | 3 | 6 | 40 | 8 | 9 |

Предложенный метод позволяет значительно увеличить длину LDPC-кода, для которого возможно точное определение кодового расстояния. Так в представленной работе было определено кодовое расстояние для тернарного LDPC-кода длиной 2400 [2400,2160,12] (см. Табл. 1), равное 12. Наибольшая длина LDPC-кода, для которого ранее определялось кодовое расстояние, равна 1369 [1369, 1224 ,10], см. ([36]).

**Заключение**

В работе предложен метод для оценки кодовых расстояний тернарных кодов GF(3) и двоичных образов недвоичных LDPC-кодов. В работе получена верхняя оценка константы масштабирующего коэффициэнта в случае приведения базиса блочным методом Коркина-Золотарева с размером блока $\beta$. Предложенный метод позволяет значительно увеличить длину LDPC-кода, для которого возможно точное определение кодового расстояния

**Список литературы**

UDK 681.391+511.9+519.612

## COMPUTING THE MINIMUM DISTANCE OF NONBINARY LDPC CODES USING BLOCK KORKIN-ZOLOTAREV METHOD

V.S. Usatjuk, PhD student of Computer Science department, SWSU (e-mail:L @Lcrypto.com)

In article present measure code distance algorithm of binary and ternary linear block code using block Korkin-Zolotarev (BKZ). Proved the upper bound on scaling constant for measure code distance of non-systematic linear block code using BKZ-method for different value of the block size. Introduced method show linear decrease of runtime from number of threads and work especially good under not dense lattices of LDPC-code. These properties allow use this algorithm to measure the minimal distance of code with length several thousand. The algorithm can further improve by transform into probabilistic algorithm using lattice enumerating pruning techniques.

Keywords: minimal distance, LDPC-code, lattice basis reduction, integer lattices, shortest vector problem, SVP, shortest basis problem, SBP, Block Korkin-Zolotarev reduction, BKZ.